\documentclass[lettersize,journal]{IEEEtran}

\usepackage{booktabs,tabularx,array,makecell}

\usepackage{amsmath,amsfonts}
\usepackage{graphicx}
\usepackage{url}
\usepackage{cite}
\usepackage{array}

\begin{document}

\title{Risk-Aware Batch Testing for Performance Regression Detection}

\author{Ali~Sayedsalehi,
        Peter~C.~Rigby,
        Gregory~Mierzwinski%
\thanks{A.~Sayedsalehi is with the Department of Computer Science and Software Engineering, Concordia University, Montreal, Quebec, Canada.}%
\thanks{P.~C.~Rigby is with the Department of Computer Science and Software Engineering, Concordia University, Montreal, Quebec, Canada.}%
\thanks{G.~Mierzwinski is with Mozilla, Potton, Quebec, Canada.}%
}

\maketitle

\begin{abstract}
Performance regression testing is essential in large-scale continuous-integration (CI) systems, yet executing full performance suites for every commit is prohibitively expensive. Prior work on performance regression prediction and batch testing has shown independent benefits, but each faces practical limitations: predictive models are rarely integrated into CI decision-making, and conventional batching strategies ignore commit-level heterogeneity.
We unify these strands by introducing a risk-aware framework that integrates machine-learned commit risk with adaptive batching. Using Mozilla Firefox as a case study, we construct a production-derived dataset of human-confirmed regressions aligned chronologically with Autoland, and fine-tune ModernBERT, CodeBERT, and LLaMA-3.1 variants to estimate commit-level performance regression risk, achieving up to 0.694 ROC-AUC with CodeBERT. The risk scores drive a family of risk-aware batching strategies (e.g., Risk-Aged Priority Batching, Risk-Adaptive Stream Batching), evaluated through realistic CI simulations. Across thousands of historical Firefox commits, our best overall configuration, Risk-Aged Priority Batching with linear aggregation (RAPB-la), yields a Pareto improvement over Mozilla's production-inspired baseline. RAPB-la reduces total test executions by 32.4\%, decreases mean feedback time by 3.8\%, maintains mean time-to-culprit at approximately the baseline level, reduces maximum time-to-culprit by 26.2\%, and corresponds to an estimated annual infrastructure cost savings of approximately \$491K under our cost model. These results demonstrate that risk-aware batch testing can reduce CI resource consumption while improving diagnostic timeliness. To support reproducibility and future research, we release a complete replication package containing all datasets, fine-tuning pipelines, and implementations of our batching algorithms.
\end{abstract}


\section{Introduction}\label{sec:intro}
\IEEEPARstart{P}{erformance} regression testing is a critical component of continuous integration (CI) in large-scale software systems such as web browsers.
For projects like Mozilla Firefox and Google Chrome, performance encompasses a wide range of user-visible and system-level metrics, including page load time, startup latency, rendering smoothness, memory consumption, and responsiveness to user input.
Unlike functional correctness, performance is an emergent property that depends on complex runtime interactions between code, compilers, hardware, and workloads, making regressions difficult to predict and diagnose.

Modern CI pipelines therefore invest heavily in performance testing, but this investment comes with substantial cost and latency.
Running full performance suites on every commit is infeasible due to long execution times, heterogeneous hardware requirements, and limited test capacity.
At Mozilla, performance regressions are monitored through systems such as Talos and Perfherder, which analyze benchmark results and raise alerts when degradations exceed predefined thresholds~\cite{mozilla-regression-policy}.
Recent reports on Mozilla show that thousands of performance alerts are generated annually, each requiring human triage and validation~\cite{mozilla-perf-dataset}.

To manage this scale, CI systems rely on selective testing practices such as batching and test selection, which reduce execution cost at the expense of delayed or coarser feedback.
In parallel, recent work has explored machine-learning-based approaches for predicting performance regressions and identifying risky changes early~\cite{beller2023meta}.
However, these two lines of work have largely evolved independently.
Prediction models are typically evaluated offline and are rarely connected to CI scheduling decisions, while batching strategies treat commits uniformly and ignore differences in regression risk.

This separation limits the practical impact of both approaches.
Without operational integration, predictive signals cannot influence when or how commits are tested, and without risk awareness, batching strategies may delay the detection of costly regressions.
Addressing this gap requires mechanisms that connect commit-level risk estimation to concrete CI testing decisions under realistic operational constraints.

\subsection{Contribution}
We address this challenge by introducing a framework that operationalizes performance regression risk within the CI testing process.
Rather than treating prediction and scheduling as separate concerns, our approach uses commit-level risk estimates directly to guide how commits are grouped and prioritized for performance testing.
This enables risk-sensitive testing decisions to be evaluated end-to-end under realistic CI conditions.

We fine-tune a transformer model on Mozilla’s real performance-regression data, derived from Perfherder~\cite{mozilla-perf-dataset} and aligned with the chronological order of the Autoland~\cite{mozilla-autoland} integration branch, to predict commit-level risk scores.  
These probabilistic risk scores are then used to guide risk-aware batch testing strategies that dynamically adapt the testing cadence and ordering to commit risk.  
High-risk commits are tested earlier and in smaller batches, while low-risk commits are deferred or grouped to reduce cost.  
This closes the loop between prediction and execution, turning commit-risk estimation into a direct driver of CI decision-making.

This framework addresses multiple long-standing challenges in prior work:
\begin{itemize}
    \item \textbf{Data realism:} Our dataset is derived directly from Mozilla’s production performance alerts, preserving real noise, class imbalance, and chronological drift.
    \item \textbf{Operational integration:} Our approach integrates model outputs into concrete CI scheduling policies, quantifying end-to-end impact on cost and latency.
    \item \textbf{Heterogeneous risk:} Conventional time- or size-based batching assumes uniform risk across commits; our risk-aware formulation adapts to heterogeneous commit risk, preventing starvation of medium-risk changes while maintaining throughput.
    \item \textbf{Scalability:} The framework scales to thousands of commits per month, requiring only static data (commit message, code diff) and no expensive dynamic profiling.
\end{itemize}

Our main contributions are as follows:
\begin{enumerate}
    \item \textbf{A Mozilla-derived dataset of performance regressions} collected from Perfherder and aligned to Autoland’s chronological order, capturing over ten thousand commits.
    \item \textbf{A transformer-based risk model} fine-tuned for structured commit messages and diffs, producing commit-level regression risk scores.
    \item \textbf{Risk-aware batching strategies}. We provide new strategies such as Risk-Adaptive Stream Batching (RASB), Risk-Aged Priority Batching (RAPB), and related variants with tunable thresholds controlling sensitivity to predicted risk.
    \item \textbf{A realistic CI simulation framework} that closely models Mozilla’s performance regression testing workflow, including batching, test selection, backfilling, build-time delays, and shared execution capacity. 
    The framework enables systematic experimentation with new testing strategies and allows their cost and latency trade-offs to be evaluated on historical Mozilla CI data.
\end{enumerate}

Our results demonstrate consistent improvements in both test-execution cost and regression detection latency, supported by a performance regression risk predictor. The best-performing risk model achieves a ROC-AUC of 0.694 (CodeBERT), indicating reliable commit-level discrimination between regression-inducing and non-regressing changes. When these risk estimates are integrated into batching, the strongest overall CI performance is obtained by Risk-Aged Priority Batching with linear aggregation (RAPB-la). Relative to Mozilla's production-inspired baseline, RAPB-la yields a Pareto improvement across all core metrics. It reduces total test executions by 32.4\%, decreases mean feedback time by 3.8\%, maintains mean time-to-culprit at approximately the baseline level, and reduces maximum time-to-culprit by 26.2\%. Mozilla reports spending approximately \$1.56M per year on infrastructure costs for the production-inspired baseline, whereas RAPB-la corresponds to an estimated annual cost of approximately \$1.07M, yielding annual savings of about \$491K. These findings show that integrating commit-level risk estimation with aging-aware batching can outperform realistic CI workflows while preserving, and in many cases improving, diagnostic timeliness.

Analyses on historical data show that performance remains robust across a range of batching parameters, and that class imbalance (typical positive rates below 1\%) can be mitigated through threshold calibration and risk-aware batching with parameter optimization.  
In practice, these improvements suggest that Mozilla’s CI pipelines could reduce machine utilization while accelerating regression detection, yielding both cost and latency benefits.
To support reproducibility and further research, we provide a complete replication package containing all datasets, model fine-tuning scripts, and our full implementations of batching algorithms~\cite{sayedsalehi2025replication}.

The remainder of the paper develops this argument from problem motivation to end-to-end evaluation. Section~\ref{sec:background} frames batch testing as the core systems problem and clarifies why conventional batching policies are insufficient for heterogeneous regression risk. Section~\ref{sec:mozilla} then grounds that problem in Mozilla’s production performance-testing workflow, showing the operational constraints and diagnostic process that motivate our design choices. Building on that setting, Section~\ref{sec:data} introduces the JIT-Mozilla-Perf dataset used to learn commit-level risk, and Section~\ref{sec:model} presents the predictor that turns commit metadata and diffs into actionable risk scores. Section~\ref{sec:simulation} then closes the loop by embedding those scores into the CI simulation framework and the proposed batching strategies. Section~\ref{sec:results} evaluates both the predictor and the batching policies to show how these components interact in terms of accuracy, latency, and cost.

\section{Batch Testing}\label{sec:background}
\noindent Batch testing refers to the practice of grouping multiple commits (or test executions) into a single testing job or “batch,” rather than testing each commit individually.
In large-scale continuous-integration (CI) systems where test executions impose high cost in time, machine usage, and infrastructure, batching can reduce total test runs, alleviate queue congestion, and improve throughput.

Empirical studies illustrate the effectiveness of batching.
For example, Beheshtian et al.~\cite{beheshtian2021batch} report that for 85\% of projects, a simple batch size of 2 (``Batch2'') or 4 (``Batch4'') achieved reductions in test executions ranging from 44\% to 50\% compared to testing every commit individually.
Feedback time also decreased by up to 37\%.
In another large-scale study analyzing 285 million test results from the Chromium project, batching achieved up to 98\% reduction in test execution time while preserving failure detection accuracy~\cite{fallahzadeh2024batching}.
Such results make batch testing a compelling strategy when per-commit testing is not sustainable.

However, conventional batch-testing strategies may exhibit limitations when applied to performance-regression settings:
\begin{itemize}
    \item Many strategies rely solely on time-based thresholds (e.g., run every hour) or size-based thresholds (e.g., batch up to $N$ commits) without accounting for heterogeneous commit risk. As a result, high-risk commits may be delayed within large batches, increasing time-to-culprit (TTC).
    \item Fixed-size or fixed-interval batching may produce overly large batches, increasing the cost of isolating the culprit when a regression occurs, or too many small batches, reducing efficiency gains, without adapting to the distribution of commit risk or test duration.
    \item Conventional approaches typically treat all commits as equal contributors to risk, ignoring that changes in critical subsystems may be far more likely to cause performance regressions than minor or low-impact modifications. As such, batch formation lacks prioritization of high-risk commits.
\end{itemize}

\section{Background on Performance Testing for Mozilla Firefox
}\label{sec:mozilla}
\noindent Mozilla maintains one of the largest open-source continuous-integration (CI) infrastructures, supporting the development of Firefox across dozens of operating systems, platforms, and hardware configurations.
Performance testing is an essential component of this pipeline, ensuring that code changes do not introduce regressions in responsiveness, rendering speed, memory use, or page-load performance.
To accurately characterize this process, we examined Mozilla’s public documentation and performance-testing data, and used the Treeherder, Phabricator, and Bugzilla APIs to extract relevant information and statistics. We iteratively refined our understanding through validation and correction of assumptions by Mozilla engineers, ensuring that both the described workflow and the derived data reflect Mozilla’s production practices.

\subsection{Treeherder and Perfherder}
The performance workflow revolves around two core systems: Treeherder~\cite{mozilla_treeherder} and Perfherder~\cite{mozilla_perfherder}.  
Treeherder is Mozilla’s unified CI dashboard that records all build and test jobs triggered by changes pushed to integration branches such as Autoland.  
Perfherder, a subsystem within Treeherder, stores structured performance measurements collected from thousands of benchmark executions (e.g., Speedometer, WebGL, responsiveness tests) across a wide matrix of configurations.

For each submitted change, a set of performance tests is executed, producing time-series metric data.  
Perfherder applies statistical techniques, typically Student’s $t$-tests and changepoint detection, to determine whether a new measurement deviates significantly from historical baselines~\cite{mozilla-perf-dataset}.  
If a degradation is detected, Perfherder generates a performance alert, linking the regressing metric to the set of commits included in the push.

\subsection{Performance Alerts and Triage}
Generated alerts flow into Mozilla’s performance sheriffing workflow~\cite{mozilla_sheriffing_workflow}.  
Performance sheriffs manually inspect alerts, validate statistical significance, correlate alerts across platforms, and ultimately identify the most likely regressing commit.  

The alert-to-culprit mapping is thus a multi-step diagnostic process involving:
\begin{itemize}
    \item filtering false positives (e.g., test noise, infrastructure anomalies);
    \item grouping related alerts across platforms, repositories, etc.;
    \item backfilling suspect commits when a batch contained multiple changes;
    \item filing performance bugs on Bugzilla once a culprit is confirmed.
\end{itemize}
This process motivates techniques that can reduce the volume of required tests and accelerate identification of regression-inducing commits.

\subsection{Batch Testing at Mozilla}
Because each push to Autoland may trigger hundreds of performance jobs, Mozilla frequently relies on batch testing, where multiple commits are bundled into a single test run.  
Batching mitigates the high infrastructure cost of executing full performance suites for every commit, but it also increases the number of suspect commits when a regression is detected.  

Batch sizes vary depending on load, platform, and release cycle, but can commonly range from a handful of commits to several dozen during peak activity.  
This creates tension between cost efficiency and diagnostic accuracy, a central motivation for risk-aware scheduling methods.

\subsection{Time Metrics}

Given a push containing one or more regression-inducing commits, time to culprit (TTC) measures how long it takes for Mozilla’s CI and sheriffing processes to surface the regressing commit.

We define three related timeliness metrics:
\begin{itemize}
    \item \textbf{Mean Feedback Time (MFT)}: the average time from when a push lands to when the first performance test result from that push becomes available. This reflects how quickly developers receive any signal (regression or not) about the performance impact of their changes.
    \item \textbf{Mean TTC}: the average time from commit landing to culprit identification across all regressing commits.
    \item \textbf{Max TTC}: the worst-case time to culprit, capturing tail latency in regression diagnosis.
\end{itemize}

Long TTC directly impacts developer productivity: regressions can propagate across dependent changes, cause wide-ranging metric noise in subsequent alerts, and increase the cost of backouts.  
Similarly, long MFT delays developers from obtaining early signals about the performance behavior of their changes, slowing feedback loops and increasing uncertainty.  
Reducing both TTC and MFT is therefore essential for performance quality and overall cycle-time efficiency in Firefox development.

\subsection{Autoland: Mozilla’s Integration Branch}
Autoland is Mozilla’s automated integration branch where most Firefox commits are initially landed.  
Developers submit patches through Phabricator, which Autoland continuously merges, builds, and tests before uplift to higher branches (e.g., mozilla-central).  
Autoland receives thousands of commits per month and triggers large-scale CI workflows on every push.

The high volume of commits, combined with expensive performance testing, makes Autoland the natural location for evaluating risk-based batching strategies.  
Its chronological structure also provides a realistic temporal ordering needed for training and evaluating predictive models that must operate under drift, imbalance, and real-time CI constraints.

\section{The JIT-Mozilla-Perf Dataset}\label{sec:data}
To fine-tune our risk prediction model and evaluate our batching strategies, we construct a Mozilla-derived dataset of performance regression--introducing commits and clean commits spanning the period from October 2024 to October 2025.
Our dataset combines information from Perfherder performance alerts, Bugzilla performance bugs, and the corresponding code changes landed on the Autoland integration branch.
We refer to this dataset as \emph{JIT-Mozilla-Perf} and release it as part of our replication package to support future research, serving as a benchmark for performance regression prediction~\cite{jit-mozilla-perf}.

\subsection{Data Extraction Pipeline}
The dataset construction follows Mozilla’s existing performance sheriffing workflow.
We begin by collecting all performance alerts generated by Perfherder during this one-year period.
Each valid regression alert links a regressing metric (e.g., Speedometer score, responsiveness time, memory consumption) to one or more commits within the corresponding Autoland push.

Performance sheriffs manually inspect alerts summaries, identify false positives, and eventually file a performance regression bug in Bugzilla once a culprit commit is confirmed.  
These bugs contain explicit references to the regression-inducing commits, determined by sheriffs via domain knowledge and the backfilling process; in practice, this process is often automated using Mozilla’s \emph{Sherlock} tool, which schedules the required backfilling jobs~\cite{mozilla_sheriffing_workflow}.  
Unlike prior work that relies on automated algorithms such as SZZ~\cite{Sliwerski2005SZZ}, our labels are derived from Mozilla’s curated human-confirmed regression reports, yielding substantially more reliable ground truth.

From each confirmed performance regression bug, we extract the culprit commit and obtain its structured diff and commit message.  
For clean commits, we collect Autoland commits that are not implicated in any performance regression bug during the same period.

\subsection{Representation}
For each performance regression sample, we construct a structured input representation that combines semantic information from the commit message with a normalized and token-efficient encoding of the code changes.

\begin{figure}[t]
    \centering
    \includegraphics[width=0.9\columnwidth]{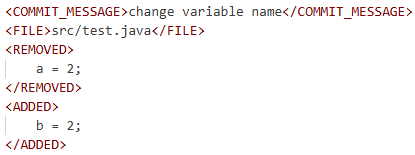}
    \caption{Example of a structured diff used as input to the ModernBERT model, reproduced from the DRS-OSS framework~\cite{sayedsalehi2025drs}. 
    Custom tokens (e.g., \texttt{<FILE>}, \texttt{<ADDED>}, \texttt{<REMOVED>}) encode the hierarchical structure of the change while reducing raw diff verbosity.}
    \label{fig:structured-diff-example}
\end{figure}

\begin{itemize}
    \item \textbf{Structured diff representation.}  
    To reduce noise in raw diffs and improve model interpretability, we transform each diff into a structured format using custom XML-style markers, following the format introduced in DRS-OSS, a defect prediction tool for code changes~\cite{sayedsalehi2025drs}.  
    Instead of passing the raw unified diff directly to the model, which is verbose, inconsistent across files, and expensive in token usage, we encode file-level and line-level changes using special tokens. A sample structured diff is shown in Figure~\ref{fig:structured-diff-example}.
    This representation compresses redundant formatting, preserves hierarchical structure, and makes the diff easier for transformer models to process.  
    The resulting structured diff includes the file paths and line-level additions and deletions of the commit.

    \item \textbf{Commit message representation.}  
    We use the commit message associated with the change.  
    The message provides semantic intent and high-level reasoning behind the modification, complementing the structural information in the diff.

    \item \textbf{Regression label.}  
    Each commit is assigned a binary label indicating whether it is regression-inducing (1) or clean (0), based on sheriff-confirmed performance regression bugs.
\end{itemize}

This representation aligns with recent findings that combining semantic intent (messages) with structural change (diff) improves predictive performance~\cite{hoang2020cc2vec}, while the structured tokenization significantly reduces input length and improves suitability for transformer-based architectures.

\subsection{Chronological Splitting and Imbalance}
Performance regressions are extremely rare.  
In our dataset, the positive class ratio is below 1\% across all splits.  
Because Mozilla’s CI operates in strict temporal order, we perform a 65/10/25 chronological split to avoid distribution leakage and mimic real deployment conditions. Table~\ref{tab:risk-dataset} reports class counts and ratios for each split. This severe imbalance, combined with naturally occurring noise and concept drift in Autoland, makes performance regression prediction a challenging modeling task.

\begin{table}[t]
\centering
\caption{Risk model dataset statistics: class counts and positive ratios in each chronological split.}
\label{tab:risk-dataset}
\begin{tabular}{lccc}
\toprule
\textbf{Split} & \textbf{Clean} & \textbf{Regressor} & \textbf{Positive \%} \\
\midrule
Train       & 7328 & 71  & 0.96\% \\
Evaluation        & 1126 & 12  & 1.05\% \\
Test  & 2832 & 15  & 0.53\% \\
\bottomrule
\end{tabular}
\end{table}

\section{Risk Model}\label{sec:model}

\noindent We frame performance regression prediction as a sequence classification problem over structured commit representations (Section~\ref{sec:data}).  
To construct our commit-level risk model, we fine-tune three transformer models, ModernBERT, CodeBERT, and LLaMA\,3.1\,8B, adapting them to output commit-level performance regression risk scores.

\subsection{Model Training and Architecture}

ModernBERT~\cite{warner2024modernbert} is a 2024 long-context encoder optimized for efficiency, extended receptive fields, and high memory throughput.  
We use the 8k-token configuration (8{,}192 max tokens), truncating longer diff sequences accordingly.  
For classification, ModernBERT passes the final hidden state of the \texttt{[CLS]} token through a linear projection followed by a softmax layer.  
Inputs are tokenized using the structured diff format described in Section~\ref{sec:data}, along with commit messages.

CodeBERT~\cite{feng2020codebert} is a bimodal encoder pretrained jointly on code and natural-language text.  
Its maximum context length is 512 tokens, so we apply aggressive truncation for longer commits.  
Like ModernBERT, CodeBERT uses the \texttt{[CLS]} pooled representation for classification, followed by a linear layer and softmax.  

LLaMA\,3.1\,8B~\cite{grattafiori2024llama3} is a decoder-only autoregressive transformer from the Llama~3 model family.  
We were able to utilize the model with contexts up to 22k tokens on a single 20\,GB NVIDIA A100 GPU by combining various efficient fine-tuning techniques:
\begin{itemize}
    \item QLoRA~\cite{dettmers2023qlora} with 4-bit NF4 quantization for base weights,
    \item LoRA adapters~\cite{Hu2021LoRA} (rank 32, $\alpha=64$),
    \item DeepSpeed ZeRO-3 CPU offloading~\cite{rajbhandari2020zeroRedundancy}
          to shard optimizer states, gradients, and parameters across CPU memory.
\end{itemize}
This combination prevents out-of-memory (OOM) errors and enables fine-tuning on a single GPU with moderate memory capacity.

For classification, we append a linear classification head to the final decoder hidden state corresponding to the last non-padding token. This “end-of-sequence embedding” acts as a pooled representation that contains information about the whole sequence.

\subsection{Output and Risk Score}
Each model outputs a probability  
\[
p(\text{regression} \mid \text{commit input}),
\]
which we interpret as a commit-level risk score.  
These scores drive the batching strategies evaluated later.

\subsection{Evaluation Metrics}

ROC-AUC measures the area under the Receiver Operating Characteristic curve, i.e., the tradeoff between true-positive rate and false-positive rate across thresholds.  
It is threshold-independent and widely used for imbalanced binary classification.

Given all commits ranked by predicted risk, recall@top-$k\%$ measures the proportion of actual regressions appearing within the highest-ranked $k\%$ of commits:
\[
\text{recall@top-}k = 
\frac{\text{\# positives in top } k\%}{\text{\# total positives}}.
\]
This metric emphasizes \emph{prioritization quality}, critical for scheduling and batching.
In Section~\ref{sec:results}, we will present the training outcomes for these metrics along with other commonly used evaluation measures.

\section{Simulation Framework}\label{sec:simulation}
\noindent This section formalizes our simulation framework and the algorithms we evaluate.
We replay the chronological stream of Autoland commits and use model-produced risk scores to drive risk-aware batch formation strategies. The simulation evaluation and test dataset splits correspond exactly to the chronological partitions used for the risk model training.

\paragraph{Inputs.}
\begin{itemize}
    \item A time-ordered commit stream $\{c_i\}$ with timestamps $t_i$.
    \item A per-commit regression probability (risk) $r_i \in [0,1]$.
    \item A ground-truth label $y_i \in \{0,1\}$ indicating whether the commit introduces a performance regression.
    \item Batching-strategy parameters (e.g., batch windows or thresholds).
    \item Empirically observed job durations extracted from Mozilla Taskcluster logs. For reference, performance test jobs have a mean runtime of 20 minutes).
\end{itemize}

\paragraph{Process.}
Commits are grouped into batches according to a chosen batching strategy and submitted for execution.
Each batch corresponds to one or more performance test jobs.
In Mozilla’s performance infrastructure, a performance test signature represents a specific performance metric measured under a particular test scenario and configuration, such as page-load time for a given benchmark on a specific platform.
In our simulation, we group multiple related performance test signatures that are executed together as a single CI job into what we refer to as a \emph{signature-group}.
This grouping reflects how performance tests are scheduled and reported as jobs in Treeherder and Perfherder, even though individual signatures are reported separately.

To reflect production behavior, we model build time as a prerequisite that must complete before any performance test job can begin. Using Mozilla Taskcluster build-time statistics for shippable builds, which are the build artifacts consumed by performance tests, we observe a mean build time of 98.7 minutes~\cite{mozilla_taskcluster} with less than 10 minutes of variation across builds. We therefore approximate build time as a constant delay in our simulation. This approximation is reasonable because both full-suite and subset-suite performance testing typically requires building the tested commit for multiple platforms, and these builds are scheduled on CI resources that are distinct from the performance-testing workers. As a result, build durations exhibit limited variability relative to the performance-testing and queueing effects that dominate overall latency. Accordingly, we apply the mean build-time delay before every performance test execution, including both batch-level runs and any subsequent backfilling runs (see subsection~\ref{subsec:backfilling}).

Test execution is modeled using a centralized test executor with a fixed pool of workers.
Each worker executes one performance test job at a time, corresponding to one signature group.
Jobs are scheduled onto the earliest available worker, allowing multiple batches and backfilling runs to overlap in time and contend for shared execution capacity.
Worker pool sizes are chosen to reflect realistic machine counts used in Mozilla’s CI infrastructure, based on worker-count statistics from Taskcluster~\cite{mozilla_taskcluster}.
We use the following platform-specific worker counts:
Android = 60,
Windows = 120,
Linux = 100,
Mac OS (includes iOS) = 250.

As a result, execution order and completion times emerge from queueing effects, resource contention, heterogeneous job durations, and build-time delays, rather than from fixed or idealized timing assumptions.

\paragraph{Outputs (Metrics).}
For each strategy we record:
total tests executed;
mean feedback time over all commits;
mean time-to-culprit (TTC) and maximum TTC over true regressors.
We perform a Pareto analysis on (total tests, max TTC) to characterize trade-offs between resource consumption and diagnostic latency.

\subsection{Batching Strategies}
Batching strategies define how consecutive commits are grouped together before performance testing.
Each strategy specifies a rule for forming batches from a chronological stream of commits, trading off testing cost against how quickly regressions are detected.
We study several families of batching strategies that range from purely time-based or size-based policies to adaptive methods that exploit predicted regression risk.

\paragraph{Time-Window Batching (TWB)}
Time-Window Batching groups commits based on when they arrive.
All commits that land within a fixed time interval (e.g., four hours) are tested together as a single batch.
This strategy bounds how long any commit can wait before being tested, but it may delay commits that arrive early in the window.
TWB treats all commits uniformly and does not account for differences in regression risk.

\paragraph{Fixed-Size Batching (FSB)}
Fixed-Size Batching groups commits into batches containing a fixed number of commits.
Once the batch size is reached, the batch is tested.
This strategy enforces a predictable number of commits per batch, but the wall-clock time covered by a batch depends on the commit arrival rate.
Like TWB, FSB ignores differences in predicted regression risk.

\paragraph{Risk-Adaptive Stream Batching (RASB)}
Risk-Adaptive Stream Batching uses predicted regression risk to decide batch boundaries.
Each commit $c_i$ is assigned a probability $r_i \in [0,1]$ that it introduces a performance regression.
Assuming independence between commits, the probability that a batch $\mathcal{B}$ is clean is
\[
P(\text{clean} \mid \mathcal{B}) = \prod_{c_i \in \mathcal{B}} (1 - r_i),
\]
and therefore the probability that the batch contains at least one regression is
\[
P(\text{fail} \mid \mathcal{B}) = 1 - \prod_{c_i \in \mathcal{B}} (1 - r_i).
\]
In the implementation, we compute this failure probability using a log-survival formulation for numerical stability.
Specifically, we accumulate the log survival mass $\sum_{c_i \in \mathcal{B}} \log(1 - r_i)$ (using a numerically stable \texttt{log1p} computation), and then recover
\[
P(\text{fail} \mid \mathcal{B}) = 1 - \exp\!\Big(\sum_{c_i \in \mathcal{B}} \log(1 - r_i)\Big).
\]
RASB appends commits to the current batch until this batch-level failure probability exceeds a chosen threshold, at which point the batch is flushed and tested.
When risk is low, large batches form, reducing testing cost.
When risk is high, batches become smaller, improving detection latency.

\paragraph{Risk-Aged Priority Batching (RAPB)}
Risk-Aged Priority Batching extends RASB by addressing fairness.
In pure risk-based strategies, commits with moderate risk may wait indefinitely if the batch-level failure probability grows too slowly.
RAPB introduces an aging mechanism that increases a commit’s effective risk the longer it waits.

For a commit $c_i$ with base risk $r_i$ and waiting time $w_i$ (in hours), RAPB computes an aged risk
\[
\tilde{r}_i = 1 - (1 - r_i)\,e^{-a w_i},
\]
where $a$ is an aging rate.
These aged risks are aggregated using the same independence model as in RASB:
\[
P(\text{fail} \mid \mathcal{B}) = 1 - \prod_{c_i \in \mathcal{B}} (1 - \tilde{r}_i).
\]
In the implementation, this batch-level failure probability is computed using the same log-survival formulation as RASB for numerical stability.
As waiting time increases, $\tilde{r}_i$ approaches 1, ensuring that batches eventually trigger testing even under prolonged low-risk conditions.
RAPB therefore balances risk sensitivity with bounded waiting time.

\paragraph{Risk-Adaptive Trigger Batching (RATB)}
Risk-Adaptive Trigger Batching combines time-based and risk-based reasoning.
Commits are grouped together as long as they remain low-risk and recent.
If a high-risk commit appears, the batch is tested immediately.
Otherwise, commits are grouped until a maximum time window is reached.
This hybrid strategy ensures prompt handling of very risky commits while bounding worst-case waiting time during low-risk periods.

\paragraph{Time-Window Subset Batching (TWSB)}
Time-Window Subset Batching models Mozilla’s production performance-testing workflow and is derived directly from historical Treeherder data.
In this workflow, different performance test signatures are executed at different frequencies, depending on platform, test cost, and scheduling policies.
As a result, each commit runs only a subset of the full performance test suite, and that subset varies over time~\cite{mozilla_sheriffing_workflow}.

This behavior effectively combines test selection and batching.
Test selection determines which performance signatures are exercised on a given commit, while batching arises implicitly because some signatures run only once every several commits or time intervals.
A performance regression therefore becomes visible only when a commit happens to execute a test signature that is sensitive to the regression.
Until then, the regression may remain undetected even though it has already been introduced.

Importantly, TWSB is not risk-aware batching.
The choice of which performance tests to run is independent of predicted commit-level regression risk and does not adapt to the content or characteristics of individual changes.
Instead, scheduling decisions are driven by static policies and resource constraints.
Because TWSB reflects how Mozilla currently balances test cost and coverage in practice, it serves as a realistic baseline for evaluating the potential benefits of explicitly risk-aware batching strategies.

\paragraph{Linear-aggregation variants (-la).}
For several risk-aware batching families, we also evaluate linear-aggregation variants, denoted by the suffix ``-la''. These variants keep the same batching and bisection structure as their corresponding base strategies, but replace probabilistic, independence-based aggregation with a simple additive risk budget. Instead of triggering a flush based on a batch-level failure probability derived from the product of $(1-r_i)$ terms, ``-la'' strategies maintain a cumulative risk sum across the current batch and flush when this sum exceeds a threshold. This linear scoring is easy to interpret and provides a direct knob for controlling the expected risk mass per batch. However, because it does not model how multiple low-risk commits combine under an independence assumption, it approximates the underlying failure likelihood more coarsely, particularly when many small risks accumulate.

\paragraph{Subset-suite variants (-s)}
For TWB, FSB, RASB, RAPB, and RATB, we also evaluate subset-suite variants, denoted by the suffix ``-s''.
These variants model Mozilla’s production test-selection behavior, similar to Time-Window Subset Batching (TWSB), where only a subset of performance test signatures is executed on each commit.
Instead of running a full performance suite for every batch, the initial batch execution uses a subset suite constructed as the union of performance signature-groups historically selected by Mozilla for the commits contained in that batch.

A regressor in a batch is detected only if the subset suite overlaps with at least one of its failing performance signature-groups.
In this way, the `-s` variants inherit the same test-selection limitations as Mozilla’s current workflow, but apply them on top of explicit batching policies.
This allows us to study how risk-aware batching strategies interact with realistic test-selection constraints, effectively combining Mozilla-style test selection with our adaptive batching methods.

\subsection{Resolution via Backfilling}
\label{subsec:backfilling}

When a batch-level performance test detects a regression, the system must identify the exact commit responsible for the degradation.
We will model the resolution process used in Mozilla’s production performance infrastructure, commonly referred to as backfilling.

\paragraph{Backfilling}
Backfilling is the process of retrospectively running targeted performance tests on individual commits within a batch after a regression has been detected.
Once a batch test signals a performance regression, the CI system schedules additional test runs for each commit contained in that batch, subject to available test capacity.
These runs execute only the relevant performance tests needed to confirm or rule out each commit as the culprit.

At Mozilla, when a performance alert is generated for a push containing multiple commits, performance sheriffs (or Sherlock) initiate backfilling jobs on Treeherder.
Each commit in the push is tested independently, often in parallel, using the same performance signatures that triggered the original alert.
As backfilling results arrive, sheriffs inspect the outcomes to determine which commit first exhibits the regression relative to known clean commits.
The earliest failing commit is then identified as the regression-inducing commit and is associated with a performance bug in Bugzilla~\cite{mozilla_sheriffing_workflow}.

Backfilling prioritizes diagnostic latency over minimizing test executions.
By exploiting parallelism in the CI infrastructure, it avoids the sequential delays inherent in iterative narrowing approaches, such as git bisect.
In our simulation, backfilling is modeled as submitting targeted tests for all commits in a failing batch concurrently and using their completion times to determine when each commit’s regression status becomes known.
This reflects Mozilla’s operational practice and provides a realistic upper bound on how quickly a batch-level regression can be resolved once detected.

\subsection{Putting It All Together}

Each experiment in our evaluation corresponds to a specific batching strategy.  
For every run, we replay the chronological commit stream, form batches using the chosen batching policy, and, upon any batch failure, invoke the backfilling algorithm to isolate regression-inducing commits.  

We report both mean feedback time and time-to-culprit (TTC) as timeliness metrics, but they capture different operational concerns in performance regression testing.
Because performance regressions are relatively rare and do not typically block development, minimizing mean feedback time is less critical in our setting: it averages over the feedback time for both clean and buggy commits, and most commits are clean.
Nevertheless, we include it for completeness and to enable comparisons with prior CI evaluation work.
In contrast, TTC matters directly once a regression is detected; rapidly identifying the regression-inducing commit is essential for mitigation and for keeping the diagnostic burden on developers and sheriffs low.

Timeliness and testing cost are inherently in tension.
We therefore focus on controlling the tail of diagnostic latency by minimizing max TTC, while simultaneously minimizing the total number of executed tests.
This cost objective is particularly important for performance testing because these jobs often run on bare metal, take longer to complete than typical unit tests, and thus consume substantial shared resources.

We follow the same chronological data-splitting methodology used in the risk model training pipeline (Section~\ref{sec:model}), dividing the data into train, eval, and test partitions.  
For each strategy combination, hyperparameters are first tuned on the \emph{eval} split, and the best configuration is then re-run \emph{unchanged} on the \emph{test} split to compute the final reported metrics.  
This mirrors standard model-selection practice and ensures that no test-set information leaks into policy optimization.


To contextualize performance, we compare against two baselines:
\begin{itemize}
    \item \emph{Exhaustive Testing (ET)}, which evaluates every commit independently by running the full performance test suite on every commit. 
    ET represents an idealized upper bound on testing coverage, but incurs prohibitive testing cost and is not feasible at scale. 
    We include ET only as a reference point to illustrate why some form of batching or selective testing is necessary for performance regression testing in large CI systems.
    \item \emph{Time-Window Subset Batching (TWSB)}, which models Mozilla’s current production performance-testing workflow. 
    Different signatures run at different frequencies, so regressions become visible only when a commit happens to execute a test that is sensitive to the regression. 
    TWSB therefore reflects the combined effect of test selection and implicit batching used in practice, but does not incorporate commit-level risk awareness.
\end{itemize}

We tune all hyperparameters using continuous multi-objective optimization with Optuna, leveraging the NSGA-II evolutionary algorithm~\cite{optuna_nsga2}. For each batching strategy, Optuna explores a continuous search space (e.g., risk thresholds, window sizes, aging coefficients) and evaluates candidate configurations on the eval split. To ensure comparable convergence across strategies with different numbers of tunable parameters, we allocate 50 Optuna optimization steps per parameter. For example, RAPB has two tunable parameters, and we therefore use 100 optimization steps for RAPB. This budgeting provides each strategy with an adequate number of iterations to identify stable, well-performing configurations. The optimization jointly minimizes two objectives: (1) total tests and (2) max TTC, capturing the trade-off between cost efficiency and isolation speed.

NSGA-II (Non-dominated Sorting Genetic Algorithm II) is a widely used algorithm for approximating Pareto frontiers.  
It maintains a diverse population of candidate solutions, ranks them using non-dominated sorting, and selects the next generation using a crowding distance heuristic that preserves diversity along the frontier.  
This makes NSGA-II well suited for our setting, where improving one objective (e.g., reducing tests) can worsen another (e.g., increasing TTC).  
The result is a set of Pareto-optimal configurations from which we select the best feasible configuration (i.e., lowest tests and better timeliness) before replaying it on the test split for final reporting.

\subsection{Infrastructure Cost Estimation}
\label{subsec:infra-cost}

Mozilla spends approximately \$390K on infrastructure for performance testing over a three-month window. We use this amount as the baseline infrastructure cost for the production-inspired TWSB strategy. For each strategy, we estimate the total infrastructure usage time consumed by test execution on bare-metal hosts from the number of test runs and their execution times, and we use that quantity to estimate infrastructure cost.

Let \(n_i\) denote the number of times test \(i\) is executed and \(t_i\) its execution time in hours on the target bare-metal infrastructure. The estimated aggregate infrastructure usage time on bare-metal hosts is computed as:
\[
T_{\mathrm{infra}} = \sum_i n_i t_i
\]

If the aggregate infrastructure usage time of the TWSB baseline is denoted by \(T_{\mathrm{infra}}^{\mathrm{TWSB}}\), then the implied unit cost for the whole infrastructure can be calculated as:
\[
c_{\mathrm{infra}} = \frac{390{,}000}{T_{\mathrm{infra}}^{\mathrm{TWSB}}}
\]

This unit cost should be interpreted as the effective cost per hour of the overall bare-metal test-runner infrastructure used for performance testing, not as the hourly cost of a single machine.

The resulting infrastructure cost is then:
\[
C_{\mathrm{infra}} = T_{\mathrm{infra}} \cdot c_{\mathrm{infra}}
\]

This formulation captures the intuition that infrastructure cost increases with both the frequency of test executions and the duration of those executions.
\section{Results and Discussion}\label{sec:results}

This section presents the empirical results of the study in two parts. We first evaluate the risk predictor, comparing the fine-tuned models on held-out regression prediction performance and discussing their practical deployment characteristics in CI settings. We then analyze the CI simulation results for the batching strategies, focusing on how different policies trade off test volume, feedback time, time-to-culprit, and infrastructure cost.

\subsection{Risk Predictor}

\noindent Table~\ref{tab:model-comparison} reports the test-set performance of the three fine-tuned predictors: ModernBERT, CodeBERT, and LLaMA~3.1~8B (QLoRA).  
Across models, overall discrimination is similar, with CodeBERT achieving the strongest ROC-AUC.  
However, the models differ in how well they prioritize the highest-risk commits. Under a strict testing budget (Recall@10\%), LLaMA and ModernBERT recover a larger fraction of regressions near the very top of the ranking, while CodeBERT lags on this early-recall measure.  
When the budget is expanded (Recall@30\%), CodeBERT improves substantially and provides the strongest recall among the three models.

These patterns are consistent with architectural trade-offs. Encoder-only architectures such as ModernBERT and CodeBERT are well suited for sequence classification because they construct bidirectional representations over the entire input and incorporate a dedicated classification embedding (the \texttt{[CLS]} token).  
Such designs have shown strong empirical performance across diverse text-classification benchmarks~\cite{devlin2019bert,liu2019roberta}.  
Decoder-only models, in contrast, rely on pooling the hidden state of the final token, which can be less directly aligned with commit-level classification objectives.  
In our setting, LLaMA's larger context window does not translate into stronger overall discrimination, but it remains competitive when prioritizing only the highest-ranked commits.

Overall, CodeBERT offers the best overall ranking quality, while LLaMA and ModernBERT provide stronger prioritization under very tight top-$k$ constraints.

\paragraph{Deployment and Inference Characteristics}

The proposed risk model is practical to deploy in standard CI environments and does not require specialized or large-scale inference infrastructure.
Each evaluated model can be hosted on a single GPU with modest VRAM, such as a 20\,GB NVIDIA A100, without model parallelism.

At inference time, commit-level risk scoring is efficient.
Even for commits with large structured diffs, end-to-end inference latency remains under 5 seconds per commit.
This latency is substantially lower than the commit submission cadence in Mozilla’s CI pipeline, where commits arrive at a granularity of minutes.
As a result, a single deployed risk model is sufficient to serve the entire Autoland commit stream, and risk scores are available almost immediately after a commit is submitted.

Fine-tuning is also computationally efficient on the same class of hardware.
By employing memory-saving techniques such as low-bit quantization, parameter-efficient adaptation, and optimizer state offloading, including the methods described in Section~\ref{sec:model}, even large models such as LLaMA\,3.1\,8B can be fine-tuned on a single 20\,GB GPU.
This enables practical model updates and retraining without requiring multi-GPU or specialized accelerator setups.

These characteristics make the risk model suitable for online use in production CI systems, where low latency, efficient retraining, and minimal infrastructure overhead are essential for integrating predictive signals into scheduling and batching decisions.

\begin{table}[t]
    \centering
    \caption{Test-set performance of fine-tuned regression-risk predictors.}
    \label{tab:model-comparison}
    \renewcommand{\arraystretch}{1.2}
        \begin{tabular}{lcccc}
            \toprule
            \textbf{Model} & \textbf{ROC-AUC} & \textbf{Recall@10\%} & \textbf{Recall@30\%} \\
            \midrule
            LLaMA-3.1\,8B  & 0.681 & 0.231 & 0.615 \\
            ModernBERT     & 0.683 & 0.23 & 0.538 \\
            CodeBERT       & 0.694 & 0.077 & 0.692 \\
            \bottomrule
        \end{tabular}
\end{table}

\subsection{Batch Testing}

This section analyzes the empirical results of our CI simulation study, focusing on how different batching strategies trade off testing cost against regression detection latency. We first contextualize the magnitude of the problem by comparing exhaustive testing with Mozilla’s production-inspired baseline. We then examine each batching family in turn, interpreting the observed metrics through the lens of each strategy’s design principles. Finally, we synthesize these results to highlight broader implications for risk-aware CI scheduling.

\paragraph{Exhaustive Testing versus Production Batching}

\begin{table}[t]
\centering
\caption{Comparison between Exhaustive Testing (ET) and Mozilla’s batching strategy (TWSB).}
\label{tab:et-vs-twsb}
\setlength{\tabcolsep}{6pt}
\renewcommand{\arraystretch}{1.1}
\begin{tabularx}{\linewidth}{>{\raggedright\arraybackslash}X c c c c}
\toprule
\textbf{Strategy} 
& \textbf{Total Tests} 
& \makecell{\textbf{Mean Feedback}\\\textbf{Time (hr)}} 
& \makecell{\textbf{Mean TTC}\\\textbf{(hr)}} 
& \makecell{\textbf{Max TTC}\\\textbf{(hr)}} \\
\midrule

TWSB
& 387,422
& 6.6
& 6.0
& 12.4 \\
\addlinespace

ET
& 12,636,342
& 4,999.3
& 3,118.0
& 6,749.2 \\

\bottomrule
\end{tabularx}
\end{table}

Table~\ref{tab:et-vs-twsb} compares Exhaustive Testing (ET) with Time-Window Subset Batching (TWSB), which models Mozilla’s current production performance-testing workflow. ET executes the full performance test suite on every commit, whereas TWSB combines implicit batching with test selection.

ET performs substantially worse than TWSB across all reported metrics. Although ET maximizes testing coverage, it generates an extreme volume of test jobs that overwhelms the available CI execution capacity. This leads to severe queueing delays, resulting in much longer feedback times and significantly delayed regression diagnosis.

In contrast, TWSB limits test volume by selectively executing subsets of performance tests over time. While this approach may delay the visibility of some regressions, it avoids infrastructure saturation and therefore delivers far faster feedback and time-to-culprit in practice.

These results demonstrate that exhaustive per-commit performance testing is not only impractical due to cost, but also counterproductive under realistic CI resource constraints. Effective CI strategies must therefore explicitly manage execution capacity through selective testing and batching rather than attempting to maximize coverage.

\paragraph{Overview of evaluated strategies and metrics.}

\begin{table*}[t]
\centering
\caption{Results comparison for batching strategies on the test split, together with estimated infrastructure cost over the 3-month simulation window. Estimated infra usage time is reported for the bare-metal test-runner infrastructure. Cost is computed using an effective infrastructure-wide rate of \$3.48 per hour of infrastructure usage. ``Cost savings'' denotes the percentage reduction in estimated cost relative to TWSB.}
\label{tab:batch-results-all}
\setlength{\tabcolsep}{7pt}
\renewcommand{\arraystretch}{1.4}
\begin{tabularx}{\textwidth}{>{\raggedright\arraybackslash}X c| c c c| c c c}
\toprule
\textbf{Batching Strategy}
& \textbf{Total Tests}
& \makecell{\textbf{Mean Feedback}\\\textbf{Time (hr)}}
& \makecell{\textbf{Mean TTC}\\\textbf{(hr)}}
& \makecell{\textbf{Max TTC}\\\textbf{(hr)}}
& \makecell{\textbf{Infra Usage Time}\\\textbf{(hr)}}
& \makecell{\textbf{Cost}\\\textbf{(\$K)}}
& \makecell{\textbf{Cost savings}\\\textbf{(\%)}} \\
\midrule

TWSB (Baseline)
& 387,422
& 6.6
& 6.0
& 12.4
& 112,108
& 390
& - \\
\addlinespace

TWB
& 317,363
& 6.5
& 6.7
& 8.7
& 93,052
& 324
& 16.9 \\
\addlinespace

TWB-s
& 259,426
& 7.0
& 7.5
& 14.1
& 76,145
& 265
& 32.1 \\
\addlinespace

FSB
& 199,156
& 9.9
& 7.9
& 13.5
& 58,370
& 203
& 47.9 \\
\addlinespace

FSB-s
& 168,796
& 9.9
& 7.9
& 13.5
& 49,614
& 173
& 55.6 \\
\addlinespace

RASB
& 275,707
& 7.9
& 6.0
& 10.5
& 80,808
& 281
& 27.9 \\
\addlinespace

RASB-s
& 187,168
& 9.0
& 7.8
& 27.5
& 54,734
& 190
& 51.3 \\
\addlinespace

RASB-la
& 275,599
& 8.1
& 5.6
& 9.6
& 80,766
& 281
& 27.9 \\
\addlinespace

RASB-la-s
& 188,023
& 8.9
& 8.2
& 27.5
& 54,991
& 191
& 51.0 \\
\addlinespace

RAPB
& 404,939
& 5.8
& 5.5
& 7.2
& 118,745
& 413
& -5.9 \\
\addlinespace

RAPB-s
& 324,559
& 6.3
& 6.2
& 14.2
& 94,712
& 330
& 15.6 \\
\addlinespace

\textbf{RAPB-la}
& \textbf{261,924}
& \textbf{6.3}
& \textbf{6.0}
& \textbf{9.2}
& \textbf{76,790}
& \textbf{267}
& \textbf{31.5} \\
\addlinespace

RAPB-la-s
& 226,302
& 7.2
& 7.1
& 23.5
& 66,461
& 231
& 40.8 \\
\addlinespace

RATB
& 303,717
& 6.9
& 6.9
& 10.8
& 89,009
& 310
& 20.5 \\
\addlinespace

RATB-s
& 227,899
& 8.2
& 9.2
& 19.6
& 66,740
& 232
& 40.5 \\

\bottomrule
\end{tabularx}
\end{table*}

Table~\ref{tab:batch-results-all} reports results on the held-out test split for the baseline (TWSB) and for each explicit batching policy.
In our application, the primary objective is to reduce total tests, because performance jobs are expensive and shared CI capacity is limited.
Subject to that cost goal, we care most about bounding diagnostic delay, captured by max TTC and mean TTC.
Mean feedback time is reported for completeness, but it is not a primary decision criterion for the reasons discussed in Section~\ref{subsec:backfilling} (``Putting It All Together'').

Table~\ref{tab:batch-results-all} also reports the estimated infrastructure cost over the three-month simulation window for each batching strategy. We use an infrastructure cost of approximately \$390K over a three-month period for the production-inspired TWSB baseline. Using the formulas in Section~\ref{subsec:infra-cost}, we divide this baseline cost by the aggregate infrastructure usage time of TWSB to derive the implied unit cost of \$3.48 per hour of infrastructure usage. We then multiply this unit cost by the infrastructure usage times reported in Table~\ref{tab:batch-results-all} to estimate the total infrastructure cost of each strategy over the three-month simulation window.

\paragraph{Baseline: TWSB as the cost--latency reference point.}
TWSB (Baseline) executes 387{,}422 tests with mean TTC 6.0 hours and max TTC 12.4 hours.
Because TWSB derives from production scheduling behavior, it implicitly mixes two effects: (i) subset execution of performance signatures and (ii) implicit batching arising from different signature frequencies.
This makes it a realistic point of comparison: strategies that reduce tests below this level are immediately attractive from an operations perspective, but only if they do not introduce unacceptable diagnostic latency.

\paragraph{Time-based batching: TWB and TWB-s.}
Time-Window Batching (TWB) reduces total tests to 317{,}363 (about an 18\% reduction relative to TWSB) while also improving max TTC to 8.7 hours.
This improvement in max TTC is consistent with TWB’s defining property: by construction, commits cannot wait longer than the configured window before being included in a batch test.
Even when a regressor occurs early in a window, it is guaranteed to be exercised at the next window flush, which limits how long the system can go without initiating backfilling for that regression.
The slight degradation in mean TTC (6.7 hours vs.\ 6.0 hours for TWSB) means that TWB does not prioritize risky commits; it simply enforces a fixed cadence.
As a result, some regressions that would have happened to be tested earlier under TWSB’s heterogeneous signature schedule may be delayed until the window boundary.
This reduction in test volume also translates into lower infrastructure cost: TWB lowers the estimated three-month infrastructure cost from \$390K for TWSB to \$324K, a savings of 16.9\%.
Thus, TWB offers a simple and predictable way to improve worst-case TTC while also reducing infrastructure usage.

The subset-suite variant TWB-s further reduces tests to 259{,}426, but its diagnostic performance degrades, especially in the tail (max TTC 14.1 hours).
This pattern aligns with how ``-s'' operate: when only a subset of signatures is executed at batch time, a regression is detectable only if the chosen subset overlaps with at least one failing signature-group.
TWB-s therefore inherits an additional source of delay that TWB does not have: even if a batch is flushed promptly, the batch’s subset suite may simply not include a sensitive signature-group, and detection (and therefore backfilling) can be deferred until a later batch whose subset happens to overlap.
Operationally, TWB is an attractive option when a team wants a simple, predictable policy that bounds worst-case TTC without needing calibrated risk scores.
TWB-s is useful when test selection is mandatory for cost reasons, but Table~\ref{tab:batch-results-all} shows that its tail risk must be accepted in exchange for the additional cost savings.

\paragraph{Size-based batching: FSB and FSB-s.}
Fixed-Size Batching (FSB) provides one of the largest cost reductions among full-suite strategies: 199{,}156 total tests, roughly half of TWSB, corresponding to estimated infrastructure cost savings of 47.9\%.
However, this cost efficiency comes with worse timeliness than TWB and the risk-aware policies: mean TTC rises to 7.9 hours and max TTC to 13.5 hours.
This trade-off is a direct consequence of FSB’s coupling to arrival rate.
When commits arrive slowly, reaching the batch-size threshold can take a long time, so a regressor may wait many hours before the first batch-level test even begins, pushing up TTC.
When commits arrive quickly, batches flush often and timeliness improves, but then test counts increase.
Because our evaluation replay includes periods of varying commit arrival rates, FSB tends to perform poorly in the worst case: the policy has no inherent wall-clock bound on waiting time.

FSB-s reduces tests further to 168{,}796 and lowers the infrastructure cost, while yielding identical TTC numbers (mean 7.9, max 13.5) in Table~\ref{tab:batch-results-all}.
This suggests that, for the chosen tuned configuration on this test split, the dominant delays are governed by the batching cadence (i.e., when the batch is flushed and backfilling is triggered), rather than by incremental differences in which signature-groups are executed at batch time.
From a practical standpoint, FSB remains a reasonable choice when an organization needs a strong and easily explainable cost cap tied to ``commits per batch,'' and when commit arrival is sufficiently steady that long waiting periods are rare (or when max TTC requirements are relaxed).
The subset-suite variant pushes the cost savings even further, but without improving timeliness, reinforcing that FSB-style strategies primarily optimize resource consumption rather than diagnostic latency. FSB-s should therefore work best in pipelines with sufficiently informative subset selection, where each flush still has a good chance of covering a failing signature-group.

\paragraph{Risk-adaptive stream batching: RASB and its variants.}
RASB (full suite) sits between TWB and FSB in cost, executing 275{,}707 tests, while matching the baseline mean TTC (6.0 hours) and keeping max TTC to 10.5 hours.
Its estimated three-month infrastructure cost is \$281K, a 27.9\% reduction relative to TWSB.
This is the key promise of risk adaptivity: when risk scores are low, RASB can aggregate many commits and reduce test count and cost; when risk scores are high, it will flush sooner, limiting diagnostic delay for likely regressors.
Compared to TWB, RASB does not enforce a hard wall-clock deadline for low-risk commits, so it can save more tests and infrastructure cost than TWB, but it also risks worse worst-case TTC if risk remains low (or is under-estimated) for too long.
The fact that max TTC remains close to baseline in Table~\ref{tab:batch-results-all} indicates that, on this workload, the risk threshold is able to trigger flushes often enough to avoid unnecessary waiting. RASB should work best in settings where the risk model can meaningfully separate quiet periods from genuinely risky bursts, so that deferred flushing mostly happens on commits that are truly unlikely to regress.

The subset-suite variant RASB-s cuts tests to 187{,}168 and lowers the infrastructure cost to \$190K, but max TTC increases sharply to 27.5 hours.
This is an important cautionary result: combining (i) risk-adaptive batching boundaries with (ii) subset-suite execution can create a worst-case scenario in which a regressor is placed into a batch (perhaps because risk was not high enough to flush immediately), and then the subset suite fails to include a sensitive signature-group, delaying detection across multiple batch flushes.
In other words, risk adaptivity can reduce the number of batch executions, which is beneficial for cost, but it also reduces the number of ``opportunities'' to run a detection-relevant signature-group when test selection is constrained.
This interaction naturally inflates the tail of TTC under ``-s'' policies, and Table~\ref{tab:batch-results-all} shows that it can be severe for RASB-s. It is therefore better suited to environments where subset construction is more accurate or broader, or where regressions usually manifest across many signatures, so that fewer batch flushes do not translate into so few viable detection opportunities.

The linear-aggregation variant RASB-la achieves nearly the same test count and cost as RASB (275{,}599 tests; \$281K over three months), while slightly improving mean TTC (5.6 hours) and improving max TTC to 9.6 hours.
This suggests that the additive risk budget provides a more effective control knob for avoiding overly large batches under sustained low-but-nonzero risk, which can happen when multiplying $(1-r_i)$ terms keeps the estimated failure probability small for longer.
In practice, RASB-la may therefore be preferred when risk scores are not well calibrated probabilistically, but still provide a useful ordering or ``risk mass'' signal.
The subset-suite linear variant (RASB-la-s) still inherits the same tail failure mode as RASB-s (max TTC 27.5 hours), indicating that aggregation choice alone does not resolve the core limitation introduced by subset selection, even though the reduced subset execution lowers the estimated cost to \$191K.

\paragraph{Risk-aged priority batching: RAPB and its variants.}
RAPB (full suite) is the strongest timeliness performer in Table~\ref{tab:batch-results-all}, with the lowest mean TTC (5.5 hours) and the lowest max TTC (7.2 hours), but it does so at the highest cost: 404{,}939 total tests and an estimated three-month infrastructure cost of \$413K, exceeding even the TWSB baseline.
This aligns with RAPB’s design goal: it explicitly prevents ``starvation'' by aging the effective risk of waiting commits upward.
In practice, aging makes the system more eager to flush, because even moderately risky commits become effectively high-risk as their waiting time grows.
That behavior improves both average and tail TTC (regressions are detected and resolved quickly), but it necessarily increases test volume and infrastructure cost because more batches are executed and backfilling is triggered more often.
RAPB is therefore appropriate when strict diagnostic service-level agreements matter more than cost; for example, during a release stabilization period, or for particularly sensitive platforms where regressions must be isolated quickly to avoid prolonged impact.

RAPB-s reduces tests to 324{,}559 and lowers the estimated cost to \$330K, closer to TWSB, but sacrifices tail TTC (max 14.2 hours).
This reflects the same ``detection opportunity'' effect observed with other ``-s'' variants: even if RAPB flushes aggressively, a regression can remain invisible if the selected subset suite does not include a regressed signature-group for that batch.
Aggressive flushing still helps mean TTC (6.2 hours), but it cannot fully protect the tail under limited coverage. RAPB-s may nevertheless work better in settings that still cannot afford full-suite execution on every flush, especially when commonly selected signatures cover most practically important regressions.

The best overall cost--latency trade-off among RAPB variants is the linear-aggregation full-suite configuration RAPB-la, which executes 261{,}924 tests while keeping mean TTC at 6.0 hours and max TTC at 9.2 hours.
Its estimated infrastructure cost is \$267K, a 31.5\% reduction relative to TWSB.
Compared to the baseline TWSB, RAPB-la reduces tests by about one third while achieving a tighter max TTC bound.
Compared to TWB-s (similar test count), RAPB-la also avoids the worst-case tail inflation induced by subset selection.
This combination of large cost reduction with controlled TTC is precisely the profile we seek for risk-aware CI scheduling under constrained resources.
Intuitively, RAPB-la benefits from two stabilizing mechanisms at once: linear aggregation prevents batches from growing too large under small-but-persistent risk, while aging ensures that commits do not wait indefinitely when the stream is quiet or risk is under-estimated.

Finally, RAPB-la-s cuts tests further to 226{,}302 and lowers the estimated three-month infrastructure cost to \$231K, but its max TTC grows to 23.5 hours.
This again highlights that subset-suite execution can dominate tail behavior: reducing tests and infrastructure cost by reducing signature coverage may save resources, but it can substantially increase worst-case diagnostic delay.
Such a policy can still be acceptable if the organization explicitly prioritizes cost (e.g., bare-metal scarcity) and is willing to tolerate that some regressions will take much longer to attribute.

\paragraph{Risk-adaptive trigger batching: RATB and RATB-s.}
RATB is a hybrid intended to combine the strengths of time bounds and risk adaptivity.
Its full-suite result (303{,}717 tests; mean TTC 6.9 hours; max TTC 10.8 hours) indicates a moderate cost reduction relative to TWSB, with an infrastructure cost of \$310K, but worse TTC than RASB-la and RAPB-la.
A plausible explanation is that the time-window constraint prevents very large batches (helping max TTC), but the policy still does not incorporate an explicit aging mechanism that forces progress when risk stays low.
Consequently, RATB can behave similarly to TWB in low-risk periods, while still being sensitive to high-risk commits, but it may flush less often than RAPB and thus achieve neither the minimal tail TTC of RAPB nor the strong cost savings of FSB. That middle-ground profile is likely to fit settings where teams want a simpler hybrid with an explicit time cap, have a highly accurate risk signal, and do not need the tighter fairness guarantees provided by aging. In particular, RATB works best when false risk spikes are rare, because spurious high-risk estimates trigger premature batch flushes and erode the policy's cost advantage.

RATB-s reduces tests to 227{,}899 and lowers the infrastructure cost, but increases mean TTC to 9.2 hours and max TTC to 19.6 hours.
As with other ``-s'' variants, limited signature coverage compounds with batching: even if the window triggers periodic flushes, detection can be postponed when the subset suite misses the failing signature-groups. RATB-s is therefore more plausible in environments where periodic flushing is still desired, and where subset selection is accurate enough that each scheduled flush is likely to exercise informative signatures.

\paragraph{Synthesis: choosing a strategy under different requirements.}
Taken together, Table~\ref{tab:batch-results-all} suggests three broad regimes.
First, if minimizing tests and infrastructure cost is the dominant requirement and elevated TTC is acceptable, FSB (and especially FSB-s) are highly cost-efficient but risk long diagnostic delays.
Second, if bounding latency is critical, aggressive policies such as RAPB provide strong latency control, but can increase infrastructure cost substantially.
Third, if the goal is to achieve a strong cost reduction while maintaining a reasonable TTC, risk-aware full-suite strategies, especially those that incorporate fairness mechanisms (aging) and robust aggregation (linear budgets), provide the most balanced Pareto trade-offs, with RAPB-la offering the clearest improvement over the production-inspired baseline on our test split.

Among all the strategies, the top balanced strategies RAPB-la, RASB-la, and RASB simultaneously reduce test volume while tightening the tail of diagnostic latency, without degrading mean TTC.
Relative to the baseline, RAPB-la reduces total tests by 32.4\% and lowers max TTC by 26.2\%, while maintaining mean TTC at baseline levels.
Notably, RAPB-la also slightly improves mean feedback time by 3.8\%, making it a Pareto improvement over the baseline across all reported metrics.
In annualized cost terms, RAPB-la yields estimated savings of about \$491K per year relative to the baseline.
Similarly, RASB-la reduces total tests by 28.86\% and lowers max TTC by 22.7\%, with a 7\% improvement in mean TTC, while yielding estimated annual savings of about \$436K.
Even without linear aggregation, RASB reduces total tests by 28.8\% and lowers max TTC by 15.2\%, while maintaining mean TTC essentially at baseline levels and yielding estimated annual savings of about \$435K.
These results indicate that risk-aware batching can deliver substantial resource savings while improving worst-case time-to-culprit, and that RAPB-la and RASB-la offer the strongest overall cost and tail-latency gains in this setting.

\section{Threats to Validity}\label{sec:threats}

\textbf{Internal validity.}
\noindent Our results depend on the correctness of Mozilla’s performance-regression labels and the fidelity of our CI simulator.
Although labels are derived from human-confirmed Bugzilla reports, some regressions or mappings may be incomplete.
The simulator also abstracts away certain operational factors, such as manual performance sheriffing steps, dynamic test worker capacity, and the fact that performance tests may sometimes need to be rerun multiple times to obtain stable measurements.
In our simulation, we assume that observed test outcomes are accurate once produced, rather than modeling repeated reruns for noise reduction.
We mitigate these threats in two ways.
First, we rely on sheriff-confirmed Bugzilla reports rather than automatically inferred labels, which improves the reliability of the ground truth.
Second, all evaluated strategies, including the Mozilla-inspired baseline, are compared under the same simulator assumptions and test-outcome model.
As a result, these simplifications may affect absolute metric values, but they are less likely to bias the comparative differences observed across batching strategies.

\textbf{Construct validity.}
We operationalize ``risk'' using predicted regression probabilities from structured commit inputs.
These signals are noisy and influenced by class imbalance and temporal drift.
Our models capture only textual changes and commit messages, omitting dynamic performance factors that may also affect regressions.
Imperfect risk estimation may therefore weaken or amplify the prioritization effects of risk-aware batching.

\textbf{External validity.}
Our study is based on Mozilla Firefox’s Autoland workflow, which features high commit volume, substantial CI scale, and extensive performance testing.
Other projects with different CI structures, test-noise characteristics, rerun practices, or alerting mechanisms may observe different trade-offs.
We mitigate this threat by designing the framework itself to be CI-agnostic: it requires only a time-ordered commit stream, commit-level risk estimates, and commit-level performance outcomes.
Even so, direct replication in other industrial or open-source settings is needed before generalizing the quantitative gains.

\textbf{Statistical conclusion validity.}
We use NSGA-II for multi-objective tuning, which introduces stochasticity and does not guarantee global optima.
Randomness in certain batching strategies can also affect TTC, and the rarity of true regressions means that results may be sensitive to a small number of cases.
We mitigate these threats by fixing seeds, evaluating all methods under identical conditions, tuning on a separate evaluation split, and replaying selected configurations on a held-out test window.
We also compare strategies using multiple complementary metrics, including total tests, mean feedback time, mean TTC, and max TTC, so that our conclusions do not depend on a single outcome measure.
\section{Literature and Discussion}\label{sec:related}

\textbf{Performance regression prediction.}
\noindent Performance regression testing has attracted growing attention, with recent work spanning test selection, risk analysis, and learning-based prediction. \emph{Perphecy}~\cite{oliveira2017perphecy} predicts whether a commit is likely to affect performance and selects a reduced subset of benchmarks to run, thereby lowering testing cost. \emph{Performance Risk Analysis (PRA)}~\cite{huang2014pra} instead uses static program analysis to estimate whether a code change is likely to introduce a performance regression, enabling test prioritization without executing the full test suite. More recent work has moved toward commit- and test-level learning. \emph{PerfJIT}~\cite{chen2022perfjit} predicts regression-prone tests for a given commit, while Meta’s \emph{SuperPerforator}~\cite{beller2023meta} shows how Transformer-based models can be deployed in practice to narrow the search space for performance-regressing changes at industrial scale. Together, these studies show the promise of predictive models for reducing the cost of performance testing, but they focus primarily on prediction accuracy or test selection rather than on how predictions should be integrated into batching policies.

\textbf{Batch testing.}
Batch testing has been widely studied as a mechanism to reduce CI resource usage. Beheshtian et al.\ \cite{beheshtian2021batch} showed that simple batching (e.g., Batch2, Batch4) yields substantial test savings across open-source projects. Fallahzadeh et al.\ \cite{fallahzadeh2024batching} compared large-scale test selection, prioritization, and batching in Chromium, finding batching to be the most effective approach for reducing overall test time. However, these methods treat all commits uniformly and do not incorporate predictive signals. Our work extends batching by introducing risk-aware selection and aging mechanisms that adapt batch size and cadence based on commit-level risk.

\textbf{Risk-aware CI and scheduling.}
Risk-guided workflows have been explored for functional testing, but rarely for performance-focused CI. Prior approaches typically integrate defect-prediction models with test selection or alert triage, but none provide a unified framework that ties risk estimation directly to batch formation. Our approach operationalizes risk at multiple layers of CI---scheduling, batch segmentation, and failure isolation---and empirically evaluates these interactions through full CI-stream simulation.

\textbf{Large language models and code-change modeling.}
Modern Transformer architectures such as BERT \cite{devlin2019bert}, RoBERTa \cite{liu2019roberta}, CodeBERT \cite{feng2020codebert}, and long-context encoders like ModernBERT \cite{warner2024modernbert} have been used in various software engineering tasks including code summarization, defect prediction, and patch classification. Recent work on QLoRA \cite{dettmers2023qlora} and ZeRO offloading \cite{rajbhandari2020zeroRedundancy} has enabled efficient fine-tuning of large models such as Llama~3 \cite{grattafiori2024llama3}. We build on these advances by combining structured diffs with a long-context encoder to model performance-regression risk in a realistic CI environment.

\textbf{Infrastructure cost estimation.}
\noindent Research on CI/CD cost--benefit analysis has emphasized that automated testing workflows should be evaluated in terms of the resources they consume, not only their speed or quality outcomes~\cite{klotins2022towards}. \emph{The Art of Testing Less without Sacrificing Quality} proposed THEO, a cost-based test selection strategy that uses historical test execution data to skip tests when their expected execution cost exceeds the expected cost of deferral. Their evaluation on large Microsoft products showed substantial reductions in test executions and corresponding cost savings while maintaining product quality guarantees~\cite{herzig2015art}. In our setting, we use a simpler and more operational cost model based on aggregate CPU consumption, which allows us to compare batching strategies directly in terms of the infrastructure resources they require.

\textbf{Positioning.}
In contrast to prior work that treats prediction and CI scheduling as separate problems, our contribution is a unified, risk-aware framework that integrates ML-based risk estimation directly into batching. This yields practical CI strategies that reduce both testing cost and regression-diagnosis latency on real-world performance data from Mozilla Firefox.

\section{Conclusion and Future Work}\label{sec:conclusion}

\noindent This work introduced a unified, risk-aware framework for performance regression testing that integrates ML-based commit risk estimation with adaptive batch testing in large-scale CI systems.
Using a production-derived dataset from Mozilla Firefox spanning over ten thousand Autoland commits and human-confirmed performance regressions, we showed that commit-level performance regression risk scores can be operationalized to directly drive batch formation and failure isolation under realistic CI constraints.
Our best-performing risk model achieves a ROC-AUC of 0.694 (CodeBERT), providing reliable discrimination between regression-inducing and non-regressing changes and enabling effective risk-guided CI decisions.
Across multiple batching families, our strongest overall configuration, Risk-Aged Priority Batching with linear aggregation (RAPB-la), yields a Pareto improvement over Mozilla's production-inspired baseline.
RAPB-la reduces total performance test executions by 32.4\%, decreases mean feedback time by 3.8\%, maintains mean time-to-culprit at approximately the baseline level, and reduces maximum time-to-culprit by 26.2\%. Mozilla reports annual infrastructure costs of approximately \$1.56M for the production-inspired baseline, whereas RAPB-la corresponds to an estimated annual cost of approximately \$1.07M, yielding annual savings of about \$491K.
Taken together, these results demonstrate that risk-aware batching can simultaneously improve cost efficiency and worst-case diagnostic timeliness, rather than trading one for the other.

Looking forward, several research directions remain open.
First, future models could augment static commit representations with dynamic signals such as coverage data, performance counters, or historical test outcomes to further refine risk estimates.
Second, deploying risk-aware batching in a live CI environment would enable validation of the observed gains under real-time queueing dynamics, hardware variability, and developer workflows.
Third, extending the framework to other large CI systems and software ecosystems would help characterize the generality of risk-aware batching under different testing regimes.
Finally, integrating risk-aware scheduling with automated triage, backout recommendation, or developer-facing explanations offers promising opportunities for building end-to-end, AI-assisted performance regression prevention pipelines.

To support reproducibility and future work, we release a complete replication package containing the dataset, model fine-tuning pipelines, and full implementations of all evaluated batching strategies~\cite{sayedsalehi2025replication}.

\section*{Acknowledgment}
The authors thank Mohamed Bilel Besbes of Concordia University for feedback on Perfherder workflows.

Portions of the manuscript text were revised with language-editing assistance from ChatGPT (OpenAI). The authors reviewed and verified all content and take full responsibility for the final manuscript.

\bibliographystyle{IEEEtran}
\bibliography{references}

\end{document}